\documentclass[aps,prb,twocolumn,groupedaddress,showpacs]{revtex4}
\usepackage{dcolumn}
\usepackage{epsfig}
\usepackage{graphicx}
\usepackage{color}

\begin{document}

\title{Titanium $\alpha - \omega$ phase transformation pathway and a predicted metastable structure}

\author{N.~A. Zarkevich,$^{1}$  and D.~D. Johnson$^{1,2}$}
\email{zarkev@ameslab.gov;   ddj@ameslab.gov}
\affiliation{$^{1}$Ames Laboratory, U.S. Department of Energy, Ames, Iowa 50011-3020 USA;}
\affiliation{$^{2}$Departments of Materials Science \& Engineering and Physics, Iowa State University, Ames, Iowa 50011-2300 USA.}

\begin{abstract}
As titanium is a highly utilized metal for structural light-weighting, its phases, transformation pathways (transition states), and structures have scientific and industrial importance.   Using a proper solid-state nudged elastic band (SS-NEB) method employing two climbing images (C2-NEB) combined with density-function theory (DFT+U) methods for accurate energetics, we detail the pressure-induced $\alpha$ (ductile) to  $\omega$ (brittle) transformation at the coexistence pressure.   We find two transition states along the minimal-enthalpy path (MEP) and discover a metastable body-centered orthorhombic (bco) structure, with stable phonons, a lower density than the endpoint phases, and decreasing stability with increasing pressure. 
\end{abstract}



\maketitle

\section{Introduction}
\begin{figure}[b]
\includegraphics[width=84mm]{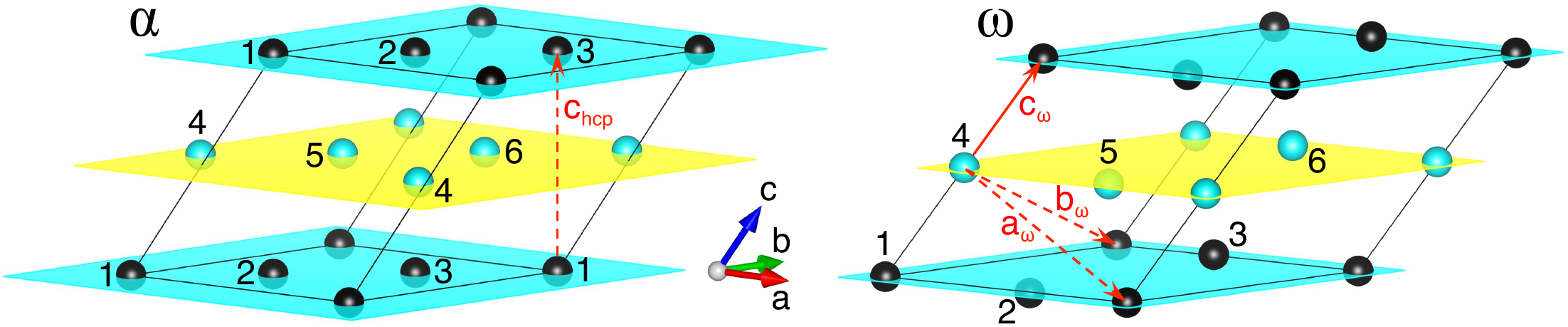}
\caption{\label{fig1str}
Enumerated 6-atom unit cells of $\alpha$ (hcp) and $\omega$ structures, suitable for the TAO-1 $\alpha-\omega$ transformation \cite{Trinkle}.
}
\end{figure}

Titanium is one of the  four (Fe, Cu, Al, Ti) most used structural metals  and is the key component of strong, lightweight structural alloys  used in aerospace, military, and automotive applications. Mapping competing phases and the associated phase transformations with stress (or pressure, $P$),  temperature ($T$), and  impurities can provide predictive design for improved control of alloy properties, including stabilizing metastable transition structures. For Ti at hydrostatic pressures above 2 GPa, the groundstate hexagonal close-packed (hcp) $\alpha$-phase can transform into a brittle higher-density $\omega$-phase \cite{Silcock,Usikov, Rabinkin} (Fig.~\ref{fig1str}).   At high $P$, Ti transforms to denser phases: $\alpha \to \omega \to \gamma \to \delta$  \cite{PRL86p3068y2001Vohra,PRL87p275503Akahama},  while at high $T$ it transforms to the body-centered cubic (bcc) $\beta$-phase \cite{PhysRev131p644y1963Jayaraman,Jamieson1963}.

{\par} Previous theoretical investigations  explored the transformation pathway -- competing structures, minimum enthalpy pathway (MEP) and transition states (TS) -- and some key results are in conflict with observations. For example, from experimental data  \cite{Zilbershteyn1973,Zilbershteyn1975,Tonkov1992,Silcock,Usikov,Rabinkin,PhysicaB355p116y2005,JPCS38p1293y1977Vohra,PRL86p3068y2001Vohra,PRL87p275503Akahama,Ming1981,JAP90p2221y2001Greeff,PhysRev131p644y1963Jayaraman,Adachi2015}, the $\alpha-\omega$  coexistence $P_0$ is $2$~GPa determined from the inequality $P_{\omega \to \alpha} < P_0 < P_{\alpha \to \omega}$  \cite{PRB91p174104}, valid for transformations between two solid anisotropic phases.
At room temperature, the $\alpha \to \omega$ transition is observed from 2-15 GPa, depending on the pressure environment and sample purity. The $\omega \to \alpha$ transformation is observed below 2 GPa \cite{Zilbershteyn1975}, but not for $P \ge 0$ for pure hydrostatic case with a gas, methanol-ethanol, or argon medium \cite{PhysicaB355p116y2005}. Deviatoric anisotropic (uniaxial or shear) stress narrows the hysteresis  \cite{PhysicaB355p116y2005,Zilbershteyn1975}.  The recent theoretical $P_0$ of 5.7 GPa \cite{JPhysChemSol69p2559y2008Zhang} disagrees with experiment \cite{Zilbershteyn1975}. 
In addition, Ti has strongly correlated $d$-electrons, and DFT returns inaccurate relative enthalpies of the groundstate and competing structures 
(e.g., hcp is not the lowest-energy structure at $0$~GPa), with a calculated $P_0 < 0$ between $\alpha-\omega$ phases \cite{Trinkle,Nmat4p129y2005}, which contradicts the experiments 
\cite{Jamieson1963,Zilbershteyn1973,Zilbershteyn1975,Tonkov1992,Silcock,Usikov,Rabinkin,PhysicaB355p116y2005,JPCS38p1293y1977Vohra,PRL86p3068y2001Vohra,PRL87p275503Akahama,Ming1981,JAP90p2221y2001Greeff,PhysRev131p644y1963Jayaraman,Adachi2015}.

{\par} Here we revisit the pressure-induced Ti $\alpha-\omega$ transformation at the coexistence pressure.  To detail the MEP and TS, we use the generalized solid-state nudged elastic band (SS-NEB) method \cite{SSNEB} based on DFT+U with onsite Hubbard corrections \cite{U} to support the required accurate relative structural enthalpies, atomic forces, and stress tensor for unit cells used for SS-NEB  \cite{SSNEB}.  Importantly, the SS-NEB method  properly couples all atomic  (or, using periodic unit cells, cell plus internal atomic) degrees of freedom and is mechanically consistent, including the MEP being invariant of cell size \cite{SSNEB}.  Adjusting $(U-J)$ to $2.2$~eV in DFT+U, we correct the inaccurate relative enthalpies and obtain the observed hcp groundstate at 0~GPa, and the observed coexistence pressure $P_0=2$ GPa; it is a value that reproduces the observed energy of reduction of Ti oxides (TiO$_2$ to Ti$_2$O$_3$), where $125~k$J/mol was matched by $(U-J)=2.3 \pm 0.1\,$eV \cite{JCTC7p2218y2011}.

{\par} Using SS-NEB combined with DFT+U, we find that the $\alpha-\omega$ transformation has two TS with a local enthalpy minimum, and discover a lower-density, body-centered orthorhombic (bco) metastable structure between them. This $\alpha \to \text{bco} \to \omega$ transformation can be considered as a sequence of two transformations. 
Impurities, pressure, and temperature control the phase stability and transition barriers in most industrial and geophysical materials -- in Ti, interstitial O, N, or C retard while substitutional Al and V suppress the $\omega$ phase \cite{Nmat4p129y2005}.
The lower-density,  bco  metastable TS structure might be stabilized by impurities or negative stresses -- potentially induced by chemical interstitial or substitutional alloying.

\section{Methods}

\begin{figure}[t]
\begin{center}
\includegraphics[width=80mm]{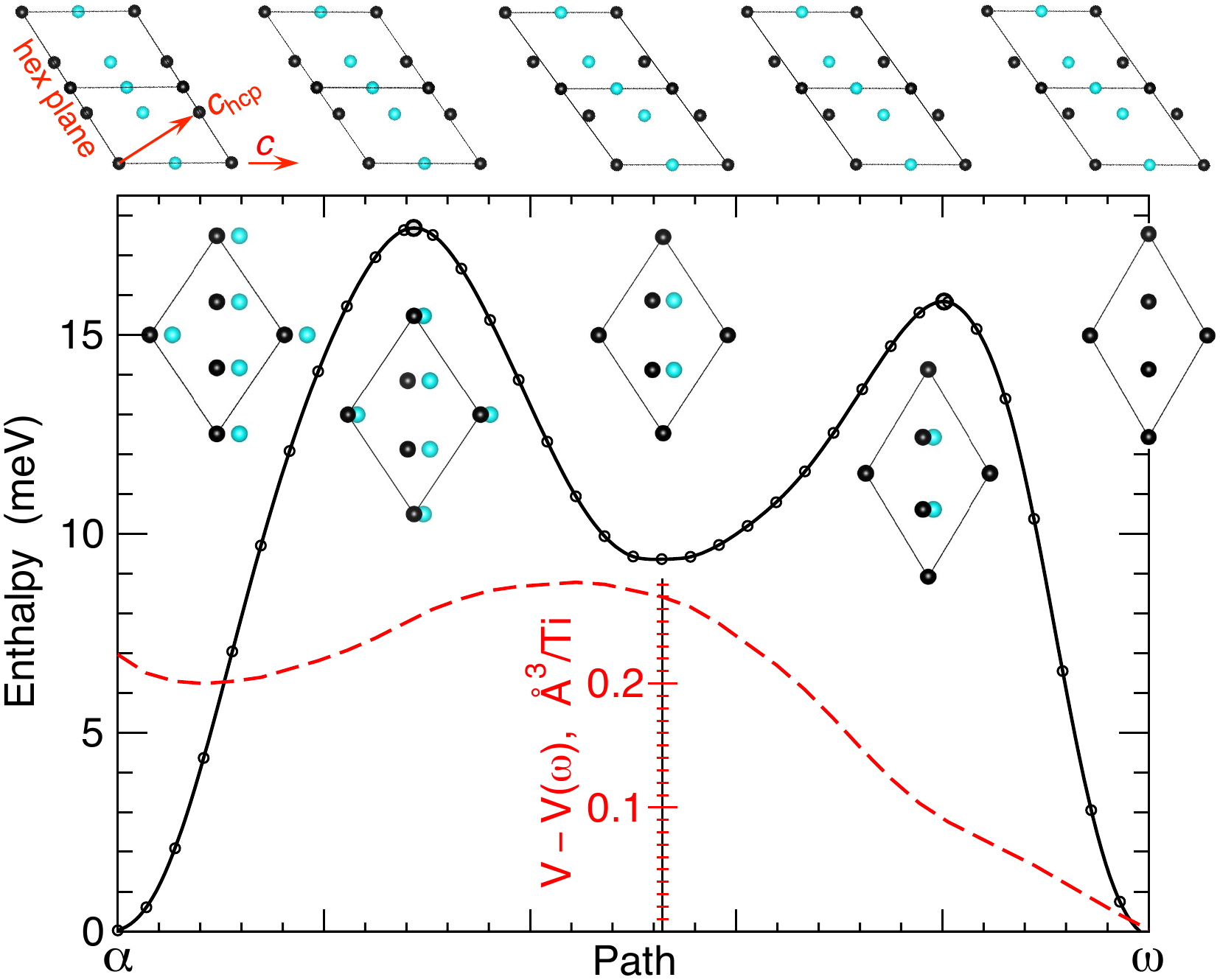}
\caption{\label{fig2GPa}
Enthalpy (meV/atom) versus MEP at $P_0=2$ GPa, where $\alpha$ and $ \omega$ enthalpies are equal within 0.15~meV/Ti. 
Dashed (red) line is volume $V$ ({\AA}$^3$/atom) relative to $\omega$ (central scale), where V($\omega$) is 17.55 {\AA}$^3$/Ti.
Atomic motion within 6-atom cell is shown for hcp $c$-axis (top): dark (black) circles and light (blue) circles indicate two hcp sub-lattices.
}
\end{center}
\end{figure}

\begin{figure}[t]
\begin{center}
\includegraphics[width=80mm]{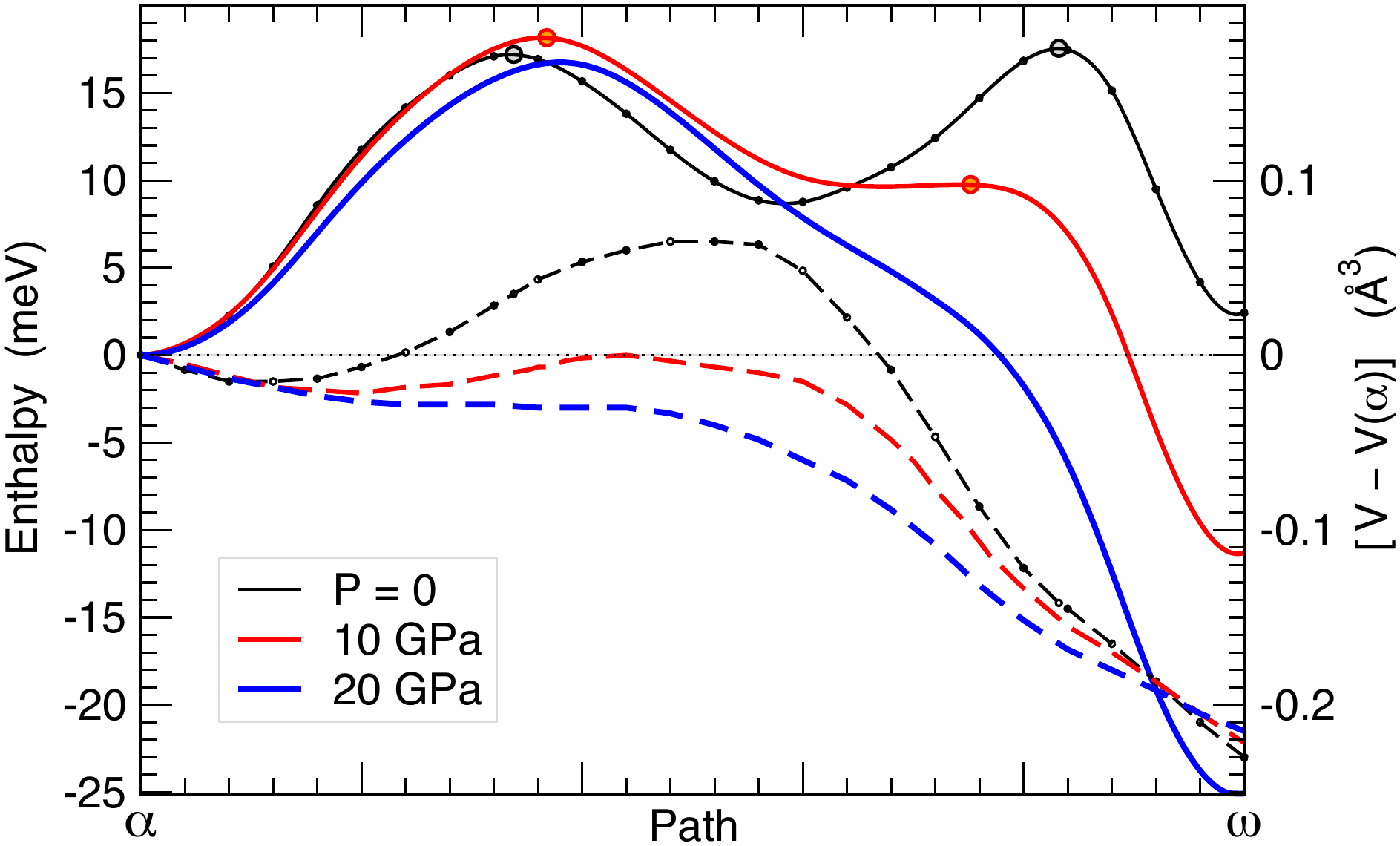}
\caption{\label{fig6MEP}
From SS-NEB (lines) and C2NEB (filled symbols), enthalpy (meV/atom) versus MEP at $P=0$, 10, and 20 GPa. Change of volume ({\AA}$^3$ per atom) relative to the $\alpha$-phase is given by dashed lines (right scale). 
}
\end{center}
\end{figure}

\begin{figure}[b]
\begin{center}
\includegraphics[width=84mm]{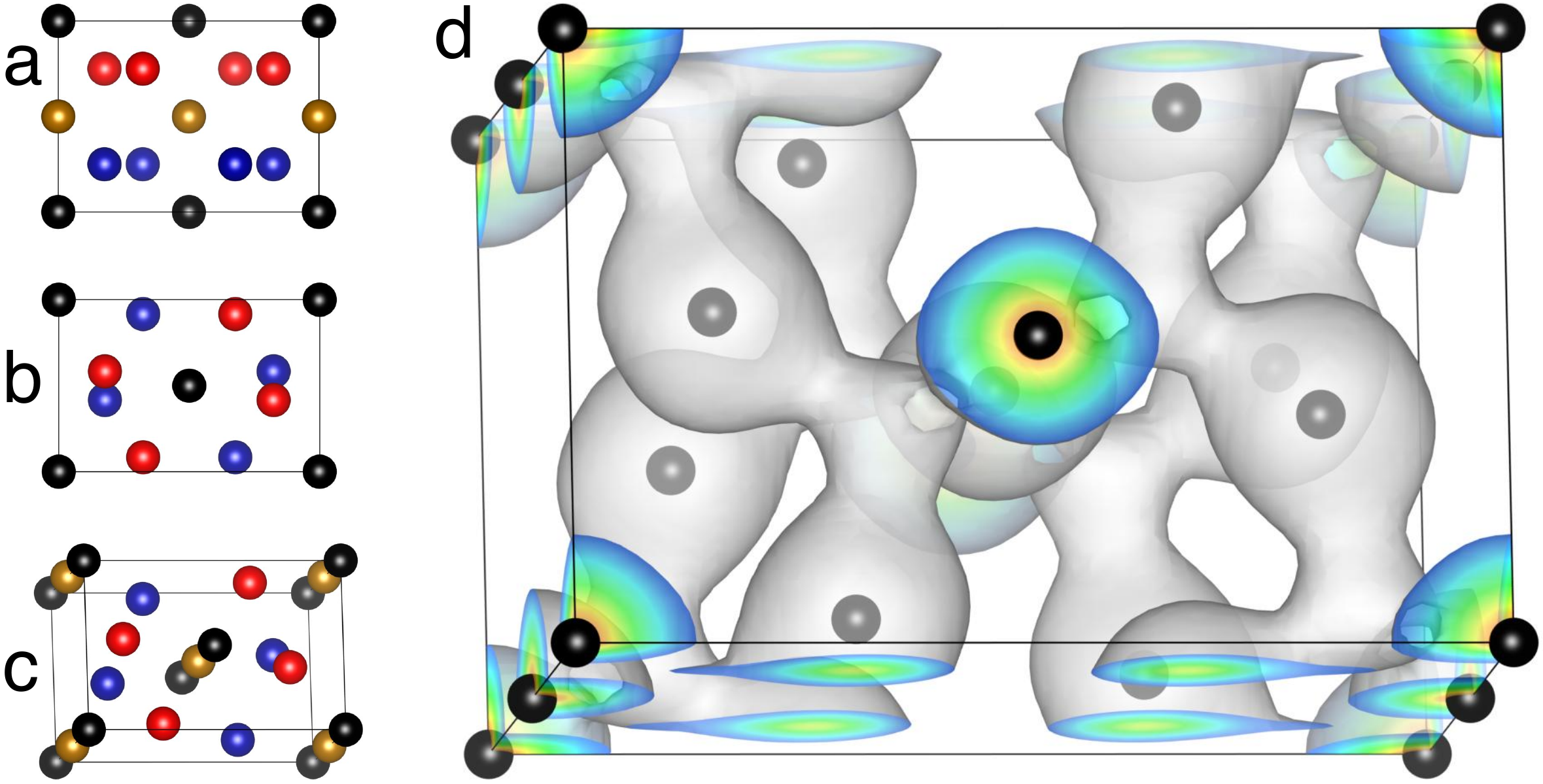}
\caption{\label{fig4bco}
12-atom (conventional) unit cell of the metastable bco structure with  layers of atoms (left), projected along $a$ (a), $b$ (b), and $c$ (c), where $a<b<c$.  (d)  iso-surfaces of electronic density ($0.033\, e^-/\mbox{\AA}^3$).
}
\end{center}
\end{figure}

The $\alpha-\omega$ transformation is considered in a 6-atom unit cell (Fig.~\ref{fig1str}). 
Applying the SS-NEB method,\cite{SSNEB} we detail the MEP (minimum enthalpy $H=E+PV$) and the transition states at coexistence pressure $P_0$ (Fig.~\ref{fig2GPa}), and versus applied pressures (Fig.~\ref{fig6MEP}). 
For accuracy, we use the C2NEB method, \cite{C2NEB,C2NEBsoft} as tested on shape-memory transforms, \cite{PRL113p265701y2014,PRB90p060102y2014} to verify each TS. 
First, we turn off climbing and then sample the path by equidistant images.  
Next, one by one, each enthalpy maximum along the path is addressed by C2NEB.  
We fully relax each local enthalpy minimum and verify its stability.  
The details of the structure and electronic density (Fig.~\ref{fig4bco}) and displacements and stress components (Fig.~\ref{fig7P}) are also provided for completeness.

{\par} We  employ DFT+U with onsite Hubbard corrections, \cite{U} as implemented in VASP \cite{VASP1,VASP2}, using projector augmented waves (PAW) \cite{PAW,PAW2}  and PW91 exchange-correlation functional. \cite{PW91}  For the 6-atom unit cell (Fig.~\ref{fig1str}), we use $12^3$ $k$-point mesh in the Brillouin zone, and a denser $24^3$ $k$-mesh for electronic density of states (DOS, see Fig.~\ref{fig3DOS}).  Gaussian smearing with $\sigma = 0.05\,$eV is used for relaxations; the tetrahedron method with Bl\"ochl corrections \cite{PRB62p6158} is used for the final total-energy calculations.
Atomic structures and data \cite{ComplexData} are visualized with VESTA \cite{Vesta3} and Grace software.\cite{xmgrace}
{\par}
Phonons for the predicted bco structure are stable (Fig.~\ref{fig5phon});  they are calculated via the small-displacement method.\cite{Phon} 
Details are given in section \ref{Results}.


\section{\label{Results}Results}
{\par} Several mechanisms for the Ti $\alpha-\omega$ transformation have been suggested.\cite{Silcock,Usikov, Rabinkin,Trinkle,Adachi2015} 
Previous DFT results \cite{Trinkle, Nmat4p129y2005} found $\omega$-phase to be Ti groundstate at 0~GPa.
In contrast, using DFT+U \cite{U} with $(U-J)$ adjusted to the experimental $P_0$ of 2~GPa -- which  matches the $(U-J)$ that also reproduces other Ti properties, such as the reduction energy of TiO$_2$  -- we obtained, not so surprisingly,  the hcp $\alpha$-Ti as the stable groundstate at  ambient pressure, in agreement with experiment.

{\par}From SS-NEB  and C2NEB calculations, we report the $\alpha - \omega$ MEP at  coexistence $P_0$ (Fig.~\ref{fig2GPa}), and MEP versus pressure (Fig.~\ref{fig6MEP}).
Clearly, we find two TS, and, in between, we find a metastable intermediate structure ($m$), which is body-centered orthorhombic (bco).  Hence, the $\alpha-\omega$ MEP consists of $\alpha-m$ and $m-\omega$ transformations, with two barriers along the $\alpha-m-\omega$ path ($18\,$meV and $16\,$meV, respectively). 
Recall that each nudged image in the SS-NEB attempts to be equidistant from its neighbors along the MEP, and minimizes its enthalpy in all other directions within the NEB code. \cite{NEB1998,SSNEB,C2NEB}
Hence, an enthalpy minimum along the MEP must be a local enthalpy minimum, i.e., a stable or metastable structure.  Indeed, being fully relaxed, the local enthalpy minimum $m$ (Fig. \ref{fig4bco}) does not transform to another structure, and, as expected, it has a stable phonon spectrum (Fig.~\ref{fig5phon}).  At low pressures, this bco structure has a lower density than the $\alpha$-phase, see volume in Figs.~\ref{fig2GPa} and \ref{fig6MEP}, and might be stabilized by dopants 
or negative stress.

\begin{figure}[t]
\begin{center}
\includegraphics[width=80mm]{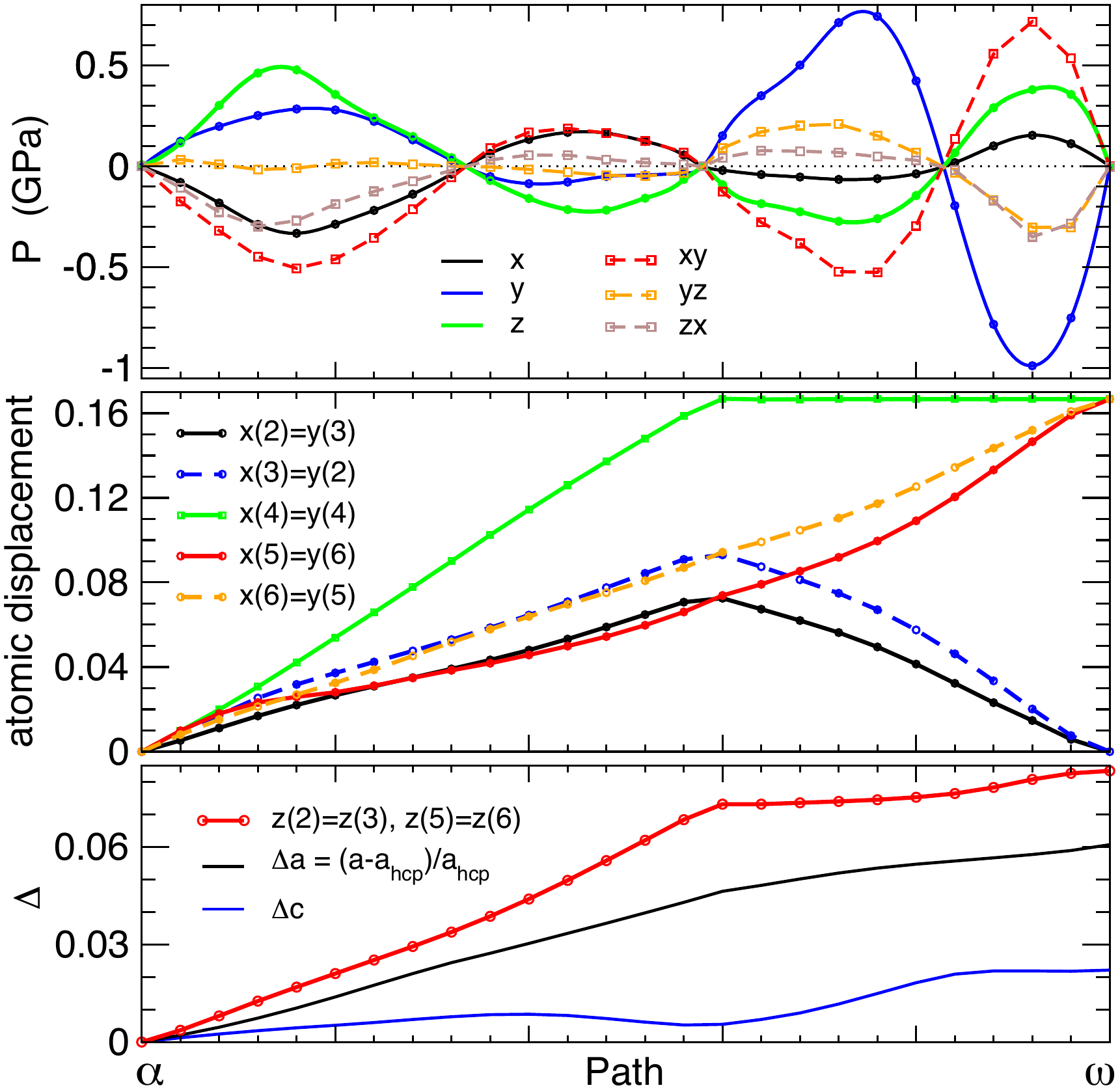}
\caption{\label{fig7P}
Diagonal $P_i$ and off-diagonal $P_{ij}$ stress components (GPa), 
absolute values of the atomic displacements $(x,y,z)$ in direct lattice coordinates (dimensionless), 
with atom 1 fixed at $(0,0,0)$, and 
elongation $\Delta$ (dimensionless) of the lattice translation vectors $a$ and $c$ (relative to the hcp $\alpha$-phase) 
in the 6-atom cell (Fig.~\ref{fig1str} and Fig.~\ref{fig2GPa} insets) 
versus MEP at zero pressure.
}
\end{center}
\end{figure}

{\par} While a new metastable bco structure is found, the MEP is still the TAO-1 (``saloon-door'' transition) path discussed by Trinkle et al.\cite{Trinkle}   Other paths, including the $\alpha$-bcc-$\omega$, suggested by Usikov \cite{Usikov} and ruled out by later experiments, \cite{PhysicaB355p116y2005} have substantially higher enthalpy barriers, in agreement with the previous calculations. \cite{Trinkle} 

{\par} At each pressure, we find two barriers in \emph{energy} $E$ for the $\alpha-\omega$ transformation. However, due to volume decrease along the MEP, the second barrier in \emph{enthalpy} $H=E+PV$ is suppressed at $P>10$ GPa, see Fig.~\ref{fig6MEP}, so the stability of the bco structure decreases with pressure. In principle, this metastable intermediate structure during the $\alpha-\omega$ transformation can be determined experimentally by x-ray diffraction (XRD), as this process might be too fast for neutron scattering.

\begin{figure}[ht]
\begin{center}
\includegraphics[width=75mm]{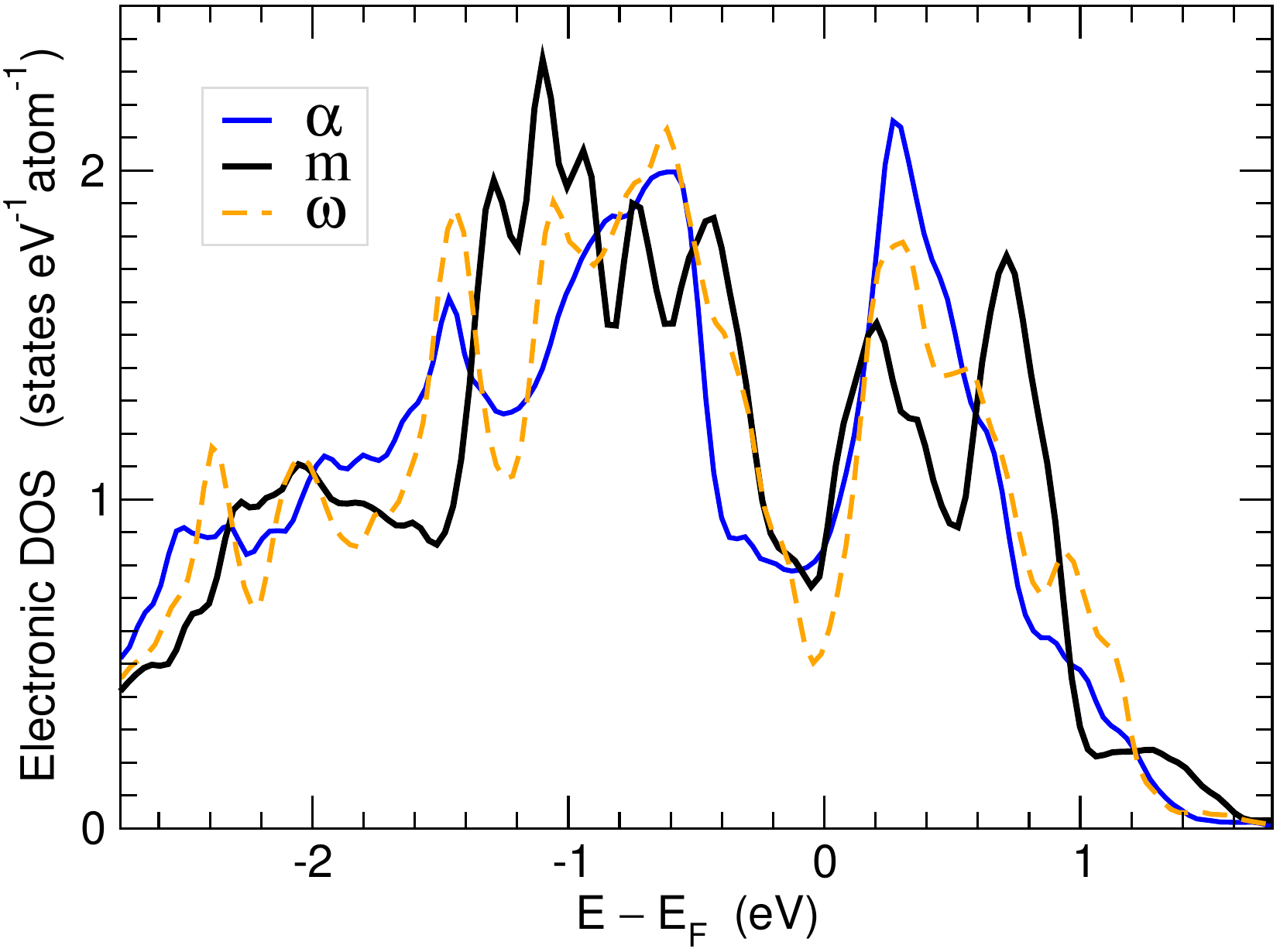}
\caption{\label{fig3DOS}
DOS versus energy $E$ (relative to $E_{F}$) for Ti $\alpha$, m, and $\omega$ phases at 2 GPa, with local minima at $E_{F}$. 
}
\end{center}
\end{figure}

{\par} Note that the transformation generates significant anisotropic stress (Fig.~\ref{fig7P}). On the other hand, pressure anisotropy can facilitate the transformation.  Indeed, an applied uniaxial or shear stress narrows the hysteresis in experiment. \cite{Zilbershteyn1975, PhysicaB355p116y2005} In fact, the reverse $\omega \to \alpha$ transformation does not happen at $P \ge 0$ under hydrostatic conditions. As expected, anisotropic stress disappears at every equilibrium point, either stable ($\alpha$, $m$, and $\omega$ structures) or unstable (both TS), see Fig.~\ref{fig7P}.
 During the transformation at $P_0$, the electronic DOS has a minimum near the Fermi energy, $E_F$, for $\alpha$, $m$, and $\omega$ structures (Fig.~\ref{fig3DOS}), as well as both TS configurations, which are the saddle points on the potential enthalpy hyper\-surface.

\begin{figure}[b]
\begin{center}
\includegraphics[width=78mm]{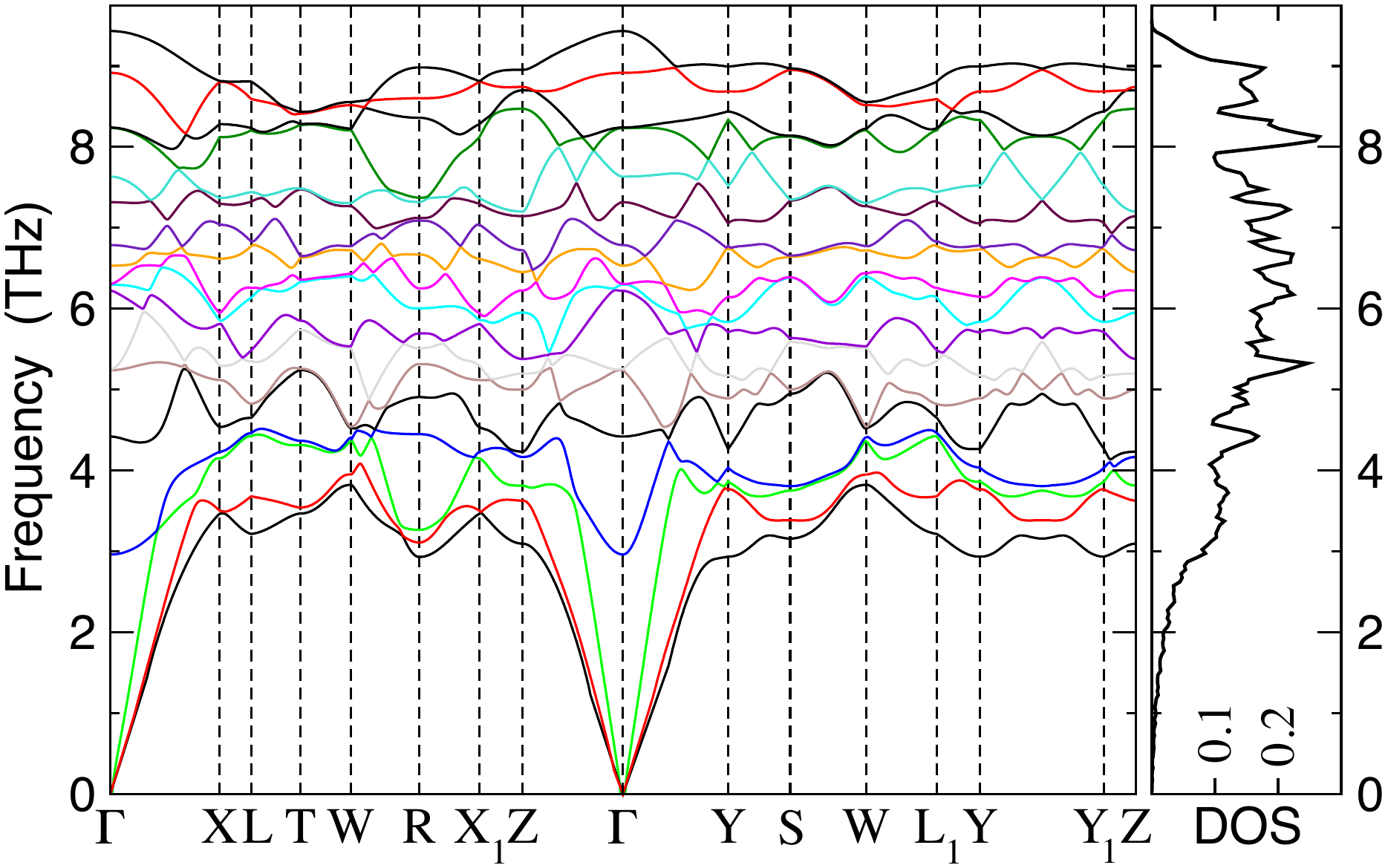}
\caption{\label{fig5phon} Phonon frequencies and DOS in the metastable bco structure at 2 GPa (coexistence pressure).}
\end{center}
\end{figure}

\section*{Phonons of  the metastable bco structure}
{\par} Phonons for the predicted bco structure are calculated via the small-displacement method, using the Phon code. \cite{Phon}  
At the $\alpha-\omega$ coexistence pressure (2~GPa), we displace each of the 6 atoms by $0.04\,${\AA} in three directions in the 162-atom $3 \times 3 \times 3$ supercell of the bco cell. The calculated atomic forces (with subtracted forces of the relaxed structure) are used to construct the force-constant matrix, symmetrized for bco.  The titanium atomic mass is $47.867~a.u.$ The phonon DOS is calculated with $0.05$~THz smearing and $21^3$ $k$-point mesh. 

{\par}
The accompanying file FORCES \cite{FORCES} provides the calculated atomic forces for each of the 18 displacements  (by $0.0025\, T_i$ along each vector $T_i$, $i=1,2,3$) of 6 atoms in the primitive unit cell (atoms 1, 28, 55, 82, 109, and 136 in the 162-atom  $3 \times 3 \times 3$ supercell,  file POSCAR. \cite{FORCES})

\section*{Structural properties}
{\par} The calculated structural parameters are given in Tables~\ref{tAtoms2GPa} and \ref{Tabc}. 
As expected, a positive Hubbard 
correction in DFT+U \cite{U} adds repulsion between electrons on the same $d$-orbital,
which results in a slight increase  of the lattice constants (which are 1\% larger than in experiment) and atomic volume $V_0$,
reported together with the bulk modulus $B_0$ and its pressure derivative $B_0^\prime$ in Table~\ref{Tabc}. 
These parameters were obtained by the least-squares fit of the Birch-Murnaghan equation of state to calculated volumes of the relaxed structures at hydrostatic pressure. 
The accuracy of the DFT+U methodology is well discussed in the literature. \cite{JCTC7p2218y2011,PRB75p035115,PRB76p155123y2007}

\begin{table}[t]
\caption{\label{tAtoms2GPa} Direct coordinates of Ti atoms in the bco structure 
in terms of the translation vectors of the primitive 6-atom unit cell,
$T_1 = (-a/2, b/2, c/2)$; 
$T_2 = ( a/2, -b/2, c/2)$;
$T_3 = ( a/2,  b/2, -c/2)$,
where the orthogonal lattice vectors at 2~GPa are
$a=5.02$; 
$b=5.58$;  
$c=7.63\,${\AA}. 
}
\begin{tabular}{lll}
   0   &     0   &     0    \\ 
   0.573337942    &    0.239952296    &    0.166666667     \\
   0.426652476    &    0.593264863    &    0.666666667     \\
   0.5    &    0    &    0.5     \\
   0.073340952   &     0.406758964    &    0.833333333     \\
   0.926655948   &     0.760071300       &     0.333333333 \\
\end{tabular}
\end{table}
\begin{table}[t]
\caption{\label{Tabc} Lattice constants ({\AA}) at $0$~GPa and Birch-Murnaghan parameters $V_0$ ({\AA}$^3$/atom), $B_0$ (GPa), and $B_0^\prime$ of the $\alpha$, $\omega$, and m phases  from DFT+U, neutron diffraction, \cite{Gray1992} and compressibility measurements.\cite{JPhysChemSolids33p1377y1972} }
\begin{tabular}{lllllllcl}
                 & $a$, $b$ & $c$ ({\AA}) & $V_0$  & $B_0$ & $B_0^\prime$ & Method \\
$\alpha$   & 2.972     & 4.728      &    18.09       &    111.7       &   3.6   & DFT+U \\
                 & 2.9506(2) & 4.6795(4) &    17.64   &    109.353  &    3.355      & Expt.  \\
$\omega$ & 4.656      & 2.854     &    17.86       & 112.9 & 3.4 & DFT+U \\
                 & 4.614(1)   & 2.832(1)   &    17.4         &           &       & Expt.  \\
m              &  5.052, 5.613   &   7.676  &      18.15     &     109.2      &   3.3    & DFT+U \\
\end{tabular}
\end{table}

\section{Summary}
We have detailed the pressure-induced Ti $\alpha-\omega$ transformation at the coexistence pressure via 
combined DFT+U \cite{U} and SS-NEB methods \cite{SSNEB,C2NEB},  using two climbing images in C2NEB \cite{C2NEB} for multiple transition states. 
With a judicious choice of $(U\!-\!J)=2.2\,$eV, DFT+U \cite{U} reproduces the observed coexistence pressure ($P_0 = 2\,$GPa) and 
 the groundstate ($\alpha$ at $P < 2\,$GPa) and provides correct relative structural enthalpies. 
It is not fortuitous that the same choice also reproduces well the reduction energies of Ti oxides \cite{JCTC7p2218y2011}. 
Importantly, we discovered a new metastable body-centered orthorhombic (bco) structure between two transition states (enthalpy barriers) along the minimal-enthalpy path.   The predicted structure has stable phonons and a lower density than the $\alpha$ and $\omega$ endpoint phases, but it has decreasing stability with increasing pressure (it is not stable above 10 GPa); 
it might be stabilized by impurities (under investigation), and provides an opportunity for engineering of lower-density titanium alloys, with additional strengthening by precipitation.

{\bf Acknowledgments: }This work was supported by the U.S. Department of Energy (DOE), Office of Science, Basic Energy Sciences, Materials Science and Engineering Division. The research was performed at the Ames Laboratory, which is operated for the U.S. DOE by Iowa State University under contract DE-AC02-07CH11358. 

\bibliography{Ti}

\begin{thebibliography}{42}
\bibitem[{FOR()}]{FORCES}
\bibinfo{note}{URL will be inserted by publisher}.
\end{thebibliography}

\end{document}